# Optimizing Service Differentiation Scheme with Sized-based Queue Management in DiffServ Networks

Nadim K. M. Madi*, and Mohamed Othman

**Abstract**— In this paper we introduced Modified Sized-based Queue Management as a dropping scheme that aims to fairly prioritize and allocate more service to VoIP traffic over bulk data like FTP as the former one usually has small packet size with less impact to the network congestion. In the same time, we want to guarantee that this prioritization is fair enough for both traffic types. On the other hand we study the total link delay over the congestive link with the attempt to alleviate this congestion as much as possible at the by function of early congestion notification. Our M-SQM scheme has been evaluated with NS2 experiments to measure the packets received from both and total link-delay for different traffic. The performance evaluation results of M-SQM have been validated and graphically compared with the performance of other three legacy AQMs (RED, RIO, and PI). It is depicted that our M-SQM outperformed these AQMs in providing QoS level of service differentiation.

**Index Terms**— QoS, AQM, Packet scheduling, Service differentiation, LIBS, M-SQM Dropping, Early Congestion Notification.

——————————— ◆ ———————————

## 1 INTRODUCTION

VARIOUS Internet-based applications are introduced daily to the world. The sharp increment of VoIP calls and the business flow worth couple billions of dollars over usual ISPs. As a result, network congestion or traffic starvation became common issue. Whereby, it turns the life over the internet pretty difficult for users. Lots of serious hiccups in the recent paradigm make the download speeds and transmission rates of heterogeneous data slow to almost dead. Over the upcoming years, Internet will no longer be able to fulfill the requirements of QoS required by current sophisticated applications. Active queue management (AQM) was introduced as a modern paradigm that copes with current Internet traffics, and deals with each of them based on its preemptive characteristics.

In [8], AQM is thought of as a pro-active approach that notifies of possible upcoming congestion situation before the buffer is flooded. Among the recent AQM schemes with the same functionality of our model is what was introduced by Kim and Yoon in [28]. Whereby, they deployed an AQM that employs multi-caching technique to identify high bandwidth flows with an attempt to provide more utilization and decrease the buffer cost as well. In [9] the authors intended to increase fairness of bandwidth allocation, and tune network capacity according to the packets precedence level. These dropping precedence are categories of packets in Per Hop Behavior PHB table.

In this paper, the prior attempt of a semilar work of service differentiation schemes in [50, 51] is extended by proposing a droping shceme with congestion notification: Modified Sized-based Queue Management (M-SQM). The principle of Less Impact Better Service (LIBS) in [35] is the key assumption of M-SQM deployment. LIBS states that: traffic flows which are not congestive and do not have significant impact to the system delay, should be prioritized over other traffic types.

M-SQM prioritizes traffic by adopting an early congestion notification mechanism. M-SQM can estimate the possibility of traffic flooding phases, and alleviate it by imposing a punishing procedure regarding a certain type of packets. According to the main purpose of M-SQM scheme wich is a service differentiation, and similar to what was mentioned in [51], we focus here in the dropping mechanism.

The dropping mechanism is deployed to decrease dropping probability of packets whose size is relatively small compared to the rest of the packets within the queue. Such a procedure occurs by employing a dynamic threshold to keep track of the average size of the incoming packets to the queue. The goal here is, to keep the value of this threshold small to some extent, so that it I will be close to the size of non-congistive packets.

In order to decrease the value of total link delay of real-time traffic, an additional congestion-aware mechanism is employed. Whereby, those packets with the big size are marked as congestive packets and replaced with smaller size packets that recently come to the queue.

Although AQM schemes are efficient to deal with heterogeneous traffic; by providing fairness, and impose level of priority to some flows according certain criterias. Nevertheless there still some gaps which are not realized, or ideally resolved yet by the prior service differentiation models. Real-time traffic is favored over the bulk data by using AQM scheme for traffic prioritization. Ironically,

————————————————

- *Corresponding author. Phone: +603 8947 1707. E-mail: nadeem.madi@gmail.com. (Nadim K M Madi). Email: mothman@fsktm.upm.edu.my. (Mohamed Othman).*
- *Nadim K M Madi and Mohamed Othman are with Department of Communication Technology and Networks, Faculty of Computer Science and Information Technology, University Putra Malaysia, 43400 UPM Serdang, Selangor, Malaysia.*

with more favoring of real-time data on the expense of FTP traffic, there is problem of starvation and significant delay impact on congestive traffic (FTP and Video traffic). Whereby, more priority is given to the small packets.

On the other hand, total link-delay is one of the main indicators for congestion situation at the congestive link between two routers. However it was not given a due care by the previous work in this domain. Rather than that, their main focus was on how to manage the queue to be favored for real-time data. The algorithm structure plays an effective role in the overall performance of the deployed model. Factors like the size of the algorithm, the number of variable and the depth of nested loops all control the efficiency. This point was not clearly guaranteed in [15, 34]. Whereby; some side-effects are counted like more memory allocation, more processing consumption, and more possibilities of fault rate are experienced.

By introducing M-SQM scheme, we are able to almost satisfy all the aforementioned problems. The issue of big traffic starvation is handled by using classification method to ensure level of fairness between different traffic types. Total link delay has been extensively considered as it is the main indicator for traffic congestion situation. A novel function is deployed in M-SQM to handle this point, by anticipating congestion stages at a time prior to the starvation. M-SQM realizes the fact of being lightweight algorithm. Utilizing only one dynamic threshold and straightforward data typing and loops makes MSQ-M outperforming its peers from the other AQMs.

The presented M-SQM scheme within this article is based on the works of [51, 52]. Wherein, service differentiation is provided for different traffic flows based on their packet sizes. The main objectives in this work are stated as following: (1) Enhance a dropping scheme that prioritizes real-time packets over FTP packets in the model of size-oriented queue management with a certain boundary. (2) Evaluate the efficiency of scheme impact within the situation of harsh congestion of traffic flows and ensure that the deployed scheme is able to detect congestion at early stages.

Following, we analyze M-SQM scheme in depth together with its impact on the traffic flow. We also evaluate it through simulation experiments that are applied to two performance metrics. The results of the simulation are then compared with another three AQM schemes from the literature. The reset of the paper is organized as following. Within section 2 we discuss the related work to our M-SQM and showing the evolution of packet classification and service differentiation. In section 3; we expose M-SQM scheme and its functionality illustrated by pseudo-code, and explain extensively about both dropping and congestion aware mechanisms. In section 4; the simulation topology and scenarios are presented, the performance metrics, as well as the experiments configuration parameters. In section 5; we show the results and discussions illustrated by graphs that compare M-SQM against the three AQM schemes. Finally, in section 6; we put on a summary to wrap up the significance of this work.

## 2 RELATED WORK

Most of service differentiation schemes work on two primitive stages: discriminate data into different traffic classes, and treat each class with various differentiation policies. Uniquely, there are two famous architectures to apply service differentiation, IntServ [42], and DiffServ [48]. IntServ was declared as standardized (RFC1 1633) in [42]. It defines the model for explaining service types and quantifying requirements of resource. Moreover, this model determines the requested resources availability which is defined at particular network components [3]. IntServ has the features of class categorization, rapid development, and matching the best-effort service class. On the other hand, DiffServ was declared as a standardized (RFC 2475) in [48]. DiffServ is provisioned-QoS model, treats the network resources as multiple classes of traffic flows with different QoS requirements [6]. The aim of DiffServ Architecture is to identify differentiated services and type-of-service (ToS) from IP version4 header, also extract the traffic class bytes from IP version 6. DiffServ enables the user to select among several of services provided and differentiated based on performance. Moreover, its traffic classes are accessible without any signaling [56]. In [59], it stated that the feature of packet classification make DiffServ architecture simpler than IntServ.

The first attempt of flow classification in packet scheduling was initiated by Sally and Van in [55] who proposed RED gateway that provides congestion avoidance in packet-switched networks, and has no bias against bursty traffic and avoids the global synchronization. However, it suffers from the inability of dynamically classify different traffic. This significant contribution was followed by the proposing of RIO [12], wherein; traffic is differentiated by tagging packets as in or out to provide estimation of TCP sending rate via designing TSW. RIO still lacks of the point that it operates with Static thresholds, TSW is unable to determine RTT of TCP, and the inability to avoid TCP retransmission timeouts. Hollot and Misra in [10] introduced PI Controller to leverage from buffer length in alleviate the queue, deliver faster responding time, and more delay control. Nevertheless, PI still requires shallow slop in the loss profile to be stable, buffer size limitation. It suffers from large queue delay, oscillatory behavior, and high loss rate in some situations.

The recent differentiated services algorithms are classified into three categories based on the indicator of traffic prioritization: (1) Size-based schemes, (2) Delay-oriented schemes, and; (3) Implicit congestion signals. Size-based scheme was introduced in SBT and SDP in [50, 51] respectively. The main determinant of class classification is the packet size. Different traffic types have different priorities depending on the average moving of packet size. A threshold is used to count for the average size of incoming packets so that preventing the queue size from being dominated by a single packet size type. However, SBT doesn't keep track of classifying control packets SDP doesn't consider the time pattern for some packets (like FTP) which require small delivery time. Our M-SQM shecme is an extensive implementation to realize the problems of the former algorithms in this category. In delay-oriented schemes like NCQ and NCQ+ in [15, 34] respectively, service differentiation procedures are performed based on the impact of each traffic type on the overall delay. However these schemes managed to prioritizes those small packets, avoid starvations from being happened, and reduces implementation and deployment efforts, their structure and complexity due to the usage of two thresholds is a negative factor restrict of being provisioned easily.

An implicit congestion signal algorithms have a unique policy of classification as introduced HtT scheme in [49]. Packets classification is done by moving certain packets from the tail of the queue to the head according to a pre-calculated probability. However the problem of explicit packet drops and extensive retransmission delays is avoided by this algorithm, it doesn't impose any solution yet to handle the real-time traffic classification. In addition to the fact that, analogy of queue rearrangement puts on more delay to the overall process.

## 3 MODIFIED SIZED-BASED QUEUE MANAGEMENT

M-SQM scheme realizes the novel QoS principle of "LIBS" which states: traffic with low impact to the overall congestion; should have more prioritized services over these congestive traffic to the queue load. M-SQM makes use of "packet size" as a criterion to apply packet classification, and early traffic congestion notification. The scheme implements two mechanisms, a dropping, in which different dropping probabilities are classified and assigned to packets, and early congestion notification which guarantee that the congestive link between the two gateways is not flooded by big sized packets.

The dropping mechanism is used to effectively differentiate incoming packets into traffic classes, and then assign various dropping probabilities toward these packets. This is usually done by defining and calculating a dynamic threshold called *msqm_thresh*; indicates the moving average of the inbound packet sizes at the queue. For each packet whose size is bigger than *msqm_thresh*; will be directly dropped with a probability same as in a classic RED gateway. Whereas, for small packets, the dropping probability is be small, and increment accordingly with the deviation ratio of the packet size from *msqm_thresh*. In this context, for purposes of obviousness; we will imply to the packets whose size is bigger than *msqm_thresh* as big packets and packets with size smaller than *msqm_thresh* as small packets. The respective flows should also be named with the same way.

The Congestion notification mechanism is more sophisticated, and works by notifying signals of expected network congestion. This is done by estimating *msqm_thresh* value and its differences from small packets. When the value of the threshold returns bigger ratios, an implicit procedure within the queue scheduler is imposed to mark the packet with relatively the biggest size as a "victim packet". This packet is revoked, removed, and replaced by a new incoming packet whose size is smaller.

### 3.1 Dropping Policy

M-SQM scheme keeps track of one dynamic variable *msqm_thresh*, which refers to the moving average of the incoming packet sizes at the router queue. For every packet arriving to the queue, *msqm_thresh* updates its value so that; it returns the most current status of the link. M-SQM scheme distinguishes packets into two types. If the packet that follows the incoming packet is greater than the latest value of the threshold, then it is called as big packets. Otherwise it is categorized as small packet.

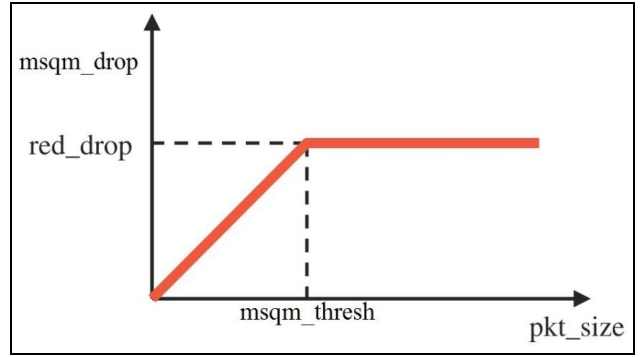

Fig. 1. MSQM dropping probability

To some extent, this classification seems to be binary, but actually there are several levels of categorization for packet sizes in the router whereby they are managed differently.

As shown in Fig. 1, M-SQM dropping probability strategy, packets whose size lies on a value beyond the threshold are dropped with a probability calculated by the original RED gateway. Small packets, on the other hand, are dropped with smaller dropping probability that is determined by its variation from *msqm_thresh*.

Let us refer to the *msqm_drop* and *red_drop* as the probabilities of M-SQM and RED respectively. *pkt_size* is the size of the incoming packet to the queue, and α is the weight factor which is equal to 0.1, then

- When *msqm_thresh* > *pkt_size* then

$$msqm\_drop = red\_drop \frac{pkt\_size}{msqm\_thresh} \qquad (1)$$

- When *msqm_thresh* ≤ *pkt_size* then

$$msqm\_drop = red\_drop \qquad (2)$$

The initial value of *msqm_thresh* before the link start to be active is the value of the first incoming packet size to the queue. The threshold is usually calculated by the value information of the weight factor from Equation (3) whereby,

$$msqm\_thresh = (1-\alpha)*msqm\_thresh + \alpha*pkt\_size \qquad (3)$$

Equations (1) and (2) illustrated clearly in the pseudo-code in Fig. 2. Packets above *msqm_thresh* are granted equivalent priority. This means that we are able to serve small packets (real-time packets) more efficiently. Nevertheless, their fair share is still confined with a level of bandwidth restriction.

**M-SQM Dropping Policy**

$pkt\_size = size(pkt)$
$msqm\_thresh = 0.9 * msqm\_thresh + 0.1 * pkt\_size$
$if\ (msqm\_thresh > pkt\_size)$
$then\ msqm\_drop = red\_drop * (pkt\_size / msqm\_thresh)$
$else\ msqm\_drop = red\_drop$
$prop = rand(0,1)$
$if\ (prop < msqm\_drop)$
$then\ drop(pkt)$
$else\ \{$
$victim = earlyCongestionNotification(pkt)$
$enqueue(pkt)$
$remove(victim)$
$pkt = victim\ \}$

Fig. 2. Pseudo-code of M-SQM scheme.

As shown, the size of the most recently arrived packets is the variable that impacts the value of *msqm_thresh*. Whereas; this threshold does not take care of the packets that have recently departed. This in turn reflects a distinct to *msqm_thresh* to be an implicit measure of the network's activity. Moreover, by granting an equivalent priority to all packet sizes greater than *msqm_thresh*, we are able to serve small packets more efficiently. However, this service is still limited with the bandwidth restriction of the fair share. The value of the weight factor α is thoroughly discussed in the next section to indicate its impact in determining the value of the threshold in M-SQM scheme.

Regarding to the rule that big packets are dropped always with a probability bigger than small packets from VoIP traffic. Our concern here is to deploy a service differentiation policy in order to guarantee a fair level of treatment to big traffic in the time that they are being restricted to a certain limit. In other words, it is not wise for us to promote so much of traffic class e.g. VoIP traffic so that the rest of traffic flows will starve. This issue is efficiently handled by M-SQM whereby; as we will demonstrate in the simulation scenarios later. There are some cases in which many small packets exist in the queue and few big packets. This phenomenon directly affects the value of *msqm_thresh* to be small and close to the small packets from VoIP traffic. Our M-SQM is featured as a light-weight algorithm due to the use of a single parameter which is combines the estimation for flows number and their packet sizes; *msqm_thresh*, as a dynamic threshold for the service differentiation. Inside the router, In order to have the most optimal estimation for this threshold, majority of packet sizes are symmetrically distributed around *msqm_thresh*.

TABLE 1
PSEUDO-CODE VARIABLES AND FUNCTIONS OF M–SQM

| Name | Description |
|---|---|
| size(pkt) | Returns the size of the packet pkt in bytes. |
| red_drop | The dropping probability of RED; computed elsewhere in the code. |
| rand(x,y) | Returns a random number between x and y. |
| enqueue(pkt) | Enqueues the packet pkt in the queue. |
| drop(pkt) | Drops the packet pkt from the queue. |
| earlyCongestionNotification(pkt) | Mark the packet as congestive packet and drop it at early stages of congestion. |
| remove(victim) | Revoke the pointer of the dropped packet from the queue slot. |

Table 1 shows more information about the interpretation of each variable and function within the M-SQM scheme body that employs both dropping and early congestion notification mechanisms.

### 3.2 Analyzing the Value of the Weight Factor α

The weight factor is a variable is with important impact on the behavior of our dynamic *msqm_thresh*. In this scope, we invoke it to show the convergence rate of *msqm_thresh* in adjusting its value fast enough to reflect the current state of the router. In this context, convergence is the ability of for *msqm_thresh* to reflect a value which is in the middle of the common different types packets sizes. According to [52], it is claimed that by setting the weight factor α = 0.1 the queue will be able to have faster convergence time. Whereby; different values of α results different convergence times. By substituting that recommended value of weight factor α at Equation (3), we have the value of *msqm_thresh* as shown in (4)

$$msqm\_thresh = 0.9 * msqm\_thresh + 0.1 * pkt\_size \quad (4)$$

Any incoming packet to the system will be classified as big packet if and only if its size is bigger than the value of *msqm_thresh*. Otherwise it is considered as small packet. Therefore, if an arrived packet to the queue determined to be big, it will be dropped early with same probability of RED *red_drop* as shown in Fig. 1.

### 3.2 Early Congestion Notification

To ensure more congestion-aware queue with less queuing link-delay, Early Congestion Notification (ECN) function with is deployed. It is assumed that, packets are confined with *msqm_thresh* authenticated to be enqueued. Authenticated here means, the packet size is smaller than the dynamic threshold. ECN guarantees more service to the real-time (small) packets, meanwhile anticipate congestion at the early stages.

As shown above in Fig. 2, ECN function picks the packet with the highest dropping probability and replaces it with an authenticated packet which has just arrived. ECN function returns the value of the marked victim packet. The victim packet here is the most congestive packet in the queue. It has relatively the biggest calculated dropping probability over other packets. Once the victim packet is removed from the queue, the authenticated packet is enqueued. Then victim packet will be hold by the procedure of "removed" to release its pointer from the queue slot. At this moment, the new enqueued packet is pointed to the queue slot of the removed victim packet. By this technique, we ensure that the queue can sustain against more burst traffic flows. The queue also can effectively perform within the congestion environment, by keep prioritizing small packets and ensure fair share of bandwidth in the same time.

## 4 PERFORMANCE EVALUATION

Performance evaluation of any computational-related system –e.g. queuing systems, operating systems, .etc, is effectively important in order to quantify the overall model Performability, such as expected throughput, system respond time, average delay, availability, reliability, and robustness. According to [38, 43], it is mentioned that; analytical modeling can be thought of as a suitable technique to evaluate straightforward queuing systems, and it comes out with more accurate results, in addition to the low cost it requires to be initiated. Nevertheless; analytical approach still requires more assumptions to be considered and this may bring a weakness point if they are not realistic enough. On the other hand, simulation has the feature of flexibility, and does not confine to any closed–formula assumptions. Therefore it is efficient in evaluating the complex queuing systems.

### 3.2 Simulation Framework

The primary objective of this model is to deploy a dropping-based scheme for classifying flows of real-time traffic, whilst keeping a level of bandwidth fair share. There is a need to implement one of the former evaluation approaches in order to precisely assess the algorithm efficiency to meet the prior objective. We believe that, in field of queuing systems, simulation is an ideal choice as an evaluation method due to the reasons delivered above. [43] Claims that, simulation technique is less restrictive in distribution of arrival rates. It has the ability to emulate queuing system behavior that cannot be detected by any other approach.

In order to perform a verification of M-SQM's that enhances QoS in service differentiation, we conduct different simulation experiments using the NS-2 Simulator. The experiments aim is twofold to satisfy the points:
1. Providing a certain level of fair share between VoIP traffic and FTP traffic in the time that VoIP traffic has the first priority of the service.
2. Control the traffic within the congestive link between the two gateways and alleviate the network traffic congestion at the early stages.

Since the initial attempts of implementing flow scheduling using RED algorithm in [30] and [55]; many service differentiation algorithms introduced to emphasize on handling heterogeneous traffic. Each traffic class is treated according to congestion notification factors –delay, queue or packet size, and drop ratio. M–SQM follows the same fundamental method in RED of classifying different traffic types. It assigns a threshold to control the queue QoS, but with more sophisticated structure supported by the principle of LIBS in [35].

The simulation framework is clearly depicted in Fig. 3. LIBS is the main principle that guides the construction of M-SQM scheme. It imposes the notion that less impact traffic types should gain better queuing services. Thereby, M-SQM formed its structure and is implemented as a complementary work to SDP, the former LIBS–based model proposed in [51]. The dropping policy in SDP has been implemented and tested in more realistic simulated Internet topology, with more user nodes which indicate more precise situation of traffic congestion.

The notion of traffic classification based on packet size is realized from the algorithm SBT in [50]. Wherein, SBT was the first attempt classifies various flows based on their packet sizes. Hereby; the former two notions (dropping policy and sized–based classification) are leveraged as core components to deploy our scheme in M-SQM. The scheme is coded in C++ language, and implanted within the RED Model in NS–2 architecture. The algorithm script was tested and debugged in NS–2 simulation over two scenarios to have more accurate outcomes. Simultaneously, we have implemented another three standard algorithms (RED, RIO, and PI) from the same domain as M–SQM in order to come out with the performance evaluation of M–SQM against these AQMs.

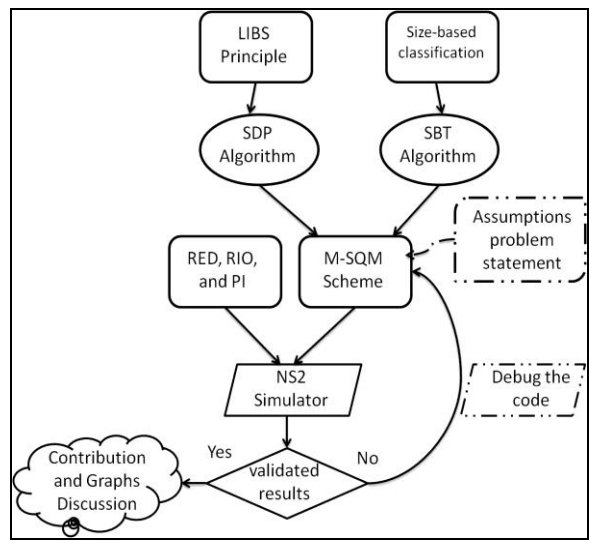

Fig. 3. MSQM Simulation freamework.

The attained statistical data of trace file was then collected and used in AWK script to measure the performance metrics "received packets" from both FTP and VoIP traffic, as well as our new metric "*Total Link–delay*" of the congestive queue. These results of M-SQM were then graphically demonstrated and analyzed with respect to the three AQMs. In case of having invalid results from M–SQM, the flow is back-tracked to debug the scheme code and execute it again with NS–2. The valid results of M–SQM are committed by graphs to reveal the significance of M-SQM scheme.

### 4.2 Performance Metrics

Due to the wide scope of this implementation; we study the performance evaluation of the model M–SQM according to two performance metrics which are "*The number of received packets*", and "*Total link delay*" for both VoIP and FTP.

**Received Packets**

Received Packets is defined as the number of successfully received packets from source nodes to the destination. However received packet seems simple and straightforward performance metric to be extracted, this parameter is influenced by many other control parameters in the

simulated network such as link delay, transmission rate, simulation time, buffer size, and the inter–transmission time (ITE) which is the time between transmitting two sequential frames or packets in FTP applications.

This parameter is calculated implicitly from a formula in NS-2 archetecture. In order to ideally guarantee the most accurate results of this parameter, several factors should be studied and numerically identified before testing and statistics collecting phases of the gained results. These factors either like model structure–based "randomness", or control parameters as mention previously.

**Total Link-delay**

Total link delay is an effective metric to estimate the congestion level within the network. The correlation between the total queuing delay and the load incrementing was considered beyond the scope of the previous related works. Therefore, we believe that the relation between the load and the total delay of the congestive link is existed. Queue length and the way it can be determined in different schemes, is one of the most effective control parameters that shape the behavior of this type of delay. Therefore it is considered carefully in this work.

Reasonably, as more number of flows (big and small packets) is pushed to the congestive link between the two routers as in Fig. 4; the queue will be flooded by many packets. At a certain point of such heavy load, this situation will cause traffic congestion. Therefore there is a need for action to alleviate this case. A noticeable ratio of queuing delay can be recorded to measure the efficiency of how fast the queue responds to eliminate this congestion.

Total link delay is calculated by an implicit formula from the trace file information. Different amount of time is captured at between the moment the packet enters the senders' gateway router, and time slice it is received by the other router.

$$Total\_link\_delay = \sum_{i=0}^{n} \left( time_{dep}[i] - time_{enq}[i] \right) \quad (5)$$

As shown in (5) above, number of flows varied from *0* to *n* enters the queue. $time_{enq}[i]$ is the array of captured time for *i* number of flows that just entered to the queue. Meanwhile, $time_{dep}[i]$ is the array of captured time for *i* number of the flows departed from the queue. Adhere; within the simulation experiments we study the behavioral impact of varied values of load on the total link delay of M–SQM. The results are compared with the other three AQM schemes.

### 4.3 AQM Traffic Types

Within M-SQM implementation scope, we used two types of traffic: (1) Big FTP traffic, and (2) VoIP Traffic. FTP traffic is carried by TCP protocol and data distribution of NewReno Version. The packet size in this traffic class is 1040 bytes (1000 bytes of FTP payload data plus 40–bytes packet header). As we mentioned previously, we call this traffic class as big packets as each packet size is greater than 500 bytes. Whereby; this size of packet seems big to the queue, and contribute more in the congestion situation.

On the other hand the VoIP traffic, are carried through UDP. Through a conversation, both speakers can alternate between OFF and ON patterns. Taking in consideration the modeling of ON and OFF patterns explained by Brady in [39], in addition to the characteristics of heavy–tailed and the feature of self–similarity traffic flows mentioned in [62], Pareto distribution is used as a data set to model the call holding times. We tuned the configuration settings of Pareto with a mean transmission rate of 78 kbps, whereas the parameter shape is set to be 1.5.

According to recommendations in [39], OFF and ON patterns are distributed with a means of 1.35 and 1.0 seconds, respectively. Based on the widely-used ITU–T G.711 coding standard proposed in [44], VoIP streams are simulated of 78 kbps with packet size set to 160 bytes (40-bytes packet header in included).

During the simulation experimental evaluation, we compared M–SQM with respect to other three AQM mechanisms: RED, PI, and RIO to measure their evaluation within two performance metrics. RED configuration sittings and parameters tuned based on the recommendation in [54]. Whereby, we implemented the "gentle" mode in RED, the maximum threshold is determined as three times the minimum threshold, and the buffer size is set to eight times the minimum threshold.

In order to have comparable results as our scheme, we use RED buffer in byte mode. However on the other hand, for RIO and PI, we used the same parameters as determined by NS–2, keeping in consideration the simulation time as well as the starting and finishing time for each of FTP and VoIP traffic flows, as they should start in the same time for the purpose of fair share.

### 4.4 Simulation Topology and Scenarios

Throughout this implementation manifest, dumbbell topology is used as shown in Fig. 4 to simulate different traffic flow types. The ultimate target of this experiment is to investigate whether M–SQM manages to allocate efficiently the network resources or result in some flows starvation. The dumbbell topology was tested over two scenarios. In the first scenario, a fixed number of 100 FTP flows sent by 100 FTP source nodes is considered. Wherein, each node generates a single flow FTP over TCP forwarded to a respective distention.

On the other hand, the VoIP traffic flows are configured to be varying from (0 – 200) flow. Whereby, each VoIP node transmits a single flow to a designated destination. Within the second scenario, we used the same topology as in the first scenario but with inverting the number of sources nodes for both FTP and VoIP. Whereby; VoIP traffic is configured to be 100 fixed flows to their designated destinations. Besides that; a varying number of FTP traffic (0 – 200) flows –each flow generated by one node.

What worth to be mentioned, according to the exhibited topology, is that the link capacity –which is low comparing to the total links capacities between the each sender and the gateway. This in turn, obligates the network to experience congestion situation by the time that more VoIP and FTP flows are transmitted to destinations. We expect to have dropping occurrences in the senders' gateway router. This means that, our queuing model was tested under the flooded situation to determine how accurate it is in dealing with traffic congistion. The link capacity of each channel is clearly shown in the topology. The capacity between each node (sender or receiver) and the respective gateway router is set to a bandwidth of 10MB and link delay of 1ms. Whereas, the congestion channel between the gateways is set to bandwidth of 50MB and link delay of 10ms.

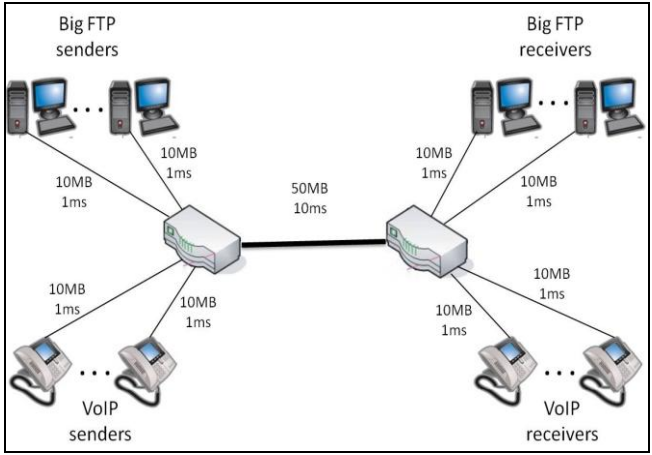

Fig. 4. MSQM Dumbbell topology.

What worth to be mentioned, according to the exhibited topology, is that the link capacity –which is low comparing to the total links capacities between the each sender and the gateway. This in turn, obligates the network to experience congestion situation by the time that more VoIP and FTP flows are transmitted to destinations. We expect to have dropping occurrences in the senders' gateway router. This means that, our queuing model was tested under the flooded situation to determine how accurate it is in dealing with traffic congistion. The link capacity of each channel is clearly shown in the topology. The capacity between each node (sender or receiver) and the respective gateway router is set to a bandwidth of 10MB and link delay of 1ms. Whereas, the congestion channel between the gateways is set to bandwidth of 50MB and link delay of 10ms.

## 5 RESULTS AND DISCUSSIONS

In the first scenario; a dumbbell topology in Fig. 4 is simulated with fixed number of 100 TCP nodes that carry FTP traffic of packet size "1040 bytes". Beside the TCP nodes, a varied number of UDP source nodes (0 – 200) have been dynamically configured in the topology. Whereby, each UDP node carries Pareto VoIP traffic forwarded to a corresponding receiver (also 0-200 nodes). Throughout this scenario the QoS behavior of each AQM scheme is studied by measuring received packets and total link-delay. The aim is to show that M-SQM is able to prioritize VoIP over FTP traffic.

In the second scenario, the same topology settings are used on M-SQM, but with inverted flow distribution among FTP and VoIP. A number of VoIP flows are maintained with a fixed number of 100. Besides that, the range of big FTP flows is gradually increased from "0 to 200".

The target here is to show that although the quota of bulk data may increase in the network M-SQM can allow fair share for buffer space between real-time and big FTP flows. Within the following we discuss the impact results of increased rate of FTP flows on both VoIP and FTP received packets. Besides that, the performance of the metric total link-delay is evaluated and discussed with varied FTP traffic.

### 5.1 First Scenario: Received Packets with Varied VoIP Traffic

The number of received packets from both traffic types FTP and VoIP have been collected from the simulation and depicted at Figures 5, and 6 respectively. First, we determined the number of FTP received packets with respect to the incrementing ratio of VoIP flows shown in Fig. 5. What worthy to be mentioned is that, the number of VoIP flows picked within the range (0-200) is not arbitrary. The reason is, any traffic load pass through the congestive link exceeds beyond 200 flows; will produce oscillatory behavior in the graph of all the competitive schemes. This is definitely expected as in that case packet loss and dropping events will be too high.

As noticed in Fig. 5, the behavior is intuitively downgraded. The main objective is to alleviate congestion situation by reducing the number of congestive FTP traffic flow. Adhere; it is clear that M-SQM yielded the best results comparing to RED, RIO, and PI. M-SQM has the lowest starting point and also the lowest end point values of the FTP received. This is because –unlike the other algorithms; M-SQM makes use of a dynamic variable *msqm_drop* that calculates the dropping probability of every incoming packet. According to the value of *msqm_drop*, ECN function is then imposed to allow the queue to mark and remove those congestive FTP packets with high probability.

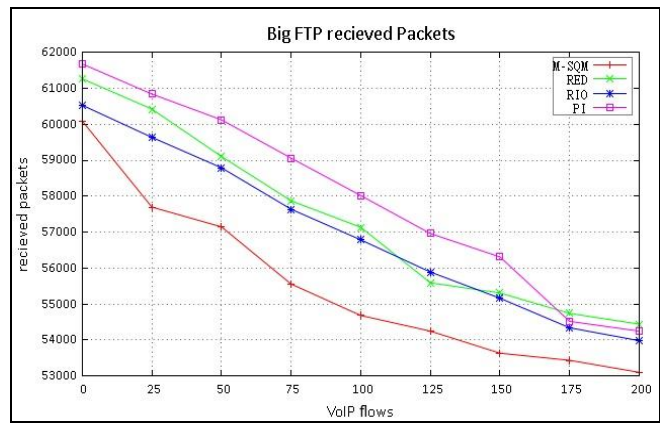

Fig. 5. FTP Received Packets with Respect to Increased VoIP Flows.

The calculated probability value ensures that the buffer will not be dominated by the big packets. This in turn, favors these small packets (VoIP traffic) over the FTP packets, and satisfies the other objective of this scope which is promoting LIBS principle.

In Fig. 6, aims to allocate more bandwidth and priority for the small VoIP traffic over FTP. As illustrated, all the algorithms record convergent results. RED looks to be slightly better at the initial stages of the flows. Nevertheless, our algorithm M–SQM is able to outperform RED at the peak points (175 – 200 flows). Nevertheless, RED cannot withstand more harsh congestion situations as more loads are increased, and here is the significance of M–SQM to outperform with its results.

Theoretically, M–SQM, aims to balance the fair share between FTP and VoIP traffic. At the initial stages of transmission, where the number of FTP flows is more than VoIP, the *msqm_thresh* definitely records higher values. This makes it a bit hard for VoIP packets to be prioritized over big FTP packets. As alternative, M–SQM provides level of balance so that FTP also can have a fair share. By the time more load rates introduced to the network, M–SQM behave differently. Wherein, there is a high expectation for traffic congestion. At this moment, the congestive FTP traffic should be compromised to cure the situation.

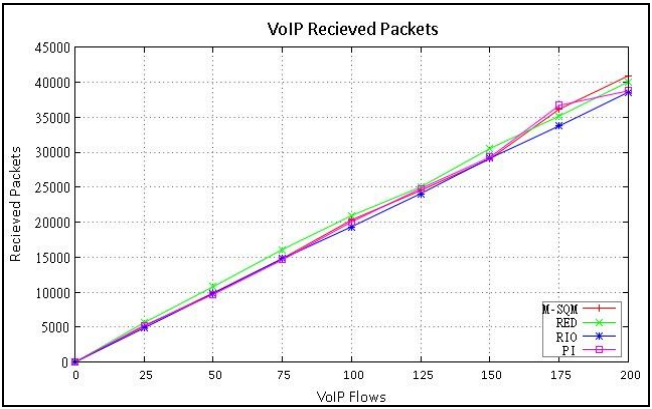

Fig. 6. VoIP received packets with respect to increased VoIP flows.

M–SQM employs the function of congestion awareness that adds another layer of filtering packets in the queue with high dropping probabilities. This is done by marking and dropping these packets; as a sign of early congestion notification. Then, the slot of the dropped packet is assign to that authorized enqueued packet. This will ensure that the queue will be maintained with light-weight size. Also, guarantee an average dropping probability near to the VoIP packet dropping value to favor more VoIP traffic as we see at the peak points.

### 5.2 First Scenario: Total Link-delay with Varied VoIP Traffic

In Fig. 7, it is clearly shown that, our deployed M–SQM outperforms the rest of the schemes by recording almost the lowest total link–delay. RED which has the nearest values to M–SQM happen to have a tiny advance in link–delay at the last points.

As more small packets come in, *msqm_thresh* become smaller. Thus, the values of dropping probability go small and too near to the random generated value. So, it is hard to mark the packets with high drop probability. Thus, add little fraction of delay link delay.

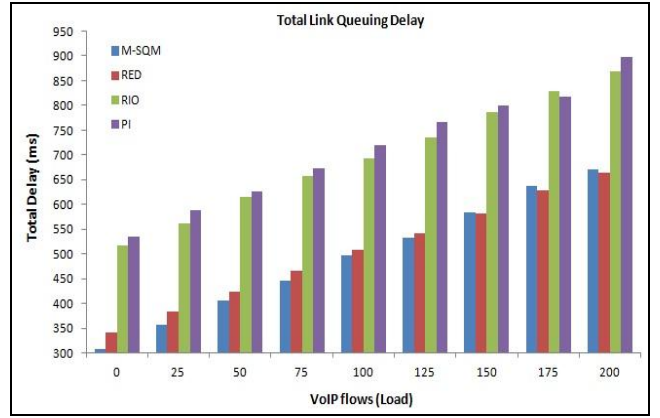

Fig. 7. Total link-delay with respect to Increased VoIP Flows.

As the VoIP load increased the queue starts to be regulated from the FTP packets and flooded with VoIP small packets. Therefore, the dropping procedure for FTP packets will be more exhausting. Whereby, the M–SQM dropping algorithm strives to calculate the biggest values of this probability and then drop its correspondence packet.

### 5.3 Second Scenario: Received Packets with Varied FTP Traffic

In Fig.8, in spite of increased number of FTP flows, M–SQM scheme yielded the lowest values of FTP received packets. This means that M–SQM outperforms its AQM peers to maintain the low rates of received FTP packets. This in turn, enables M–SQM is to rapidly regulate the queue from congestive traffic, and guarantee more chances for real-time data (VoIP packets) are ensured to be transmitted.

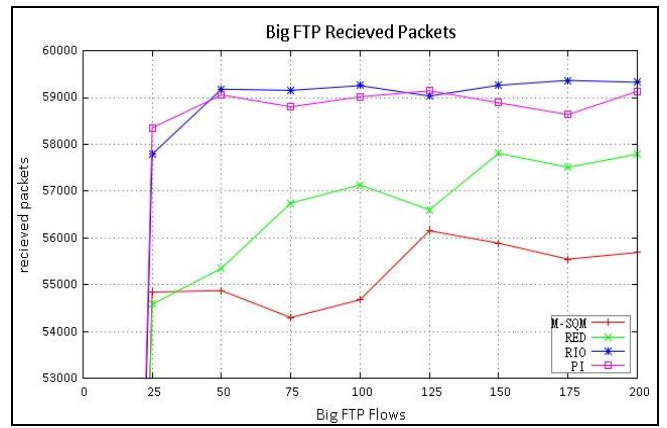

Fig. 8. FTP received packets with respect to increased FTP flows.

The initial phases of M–SQM behavior look steady. With more introduced FTP flows, the M–SQM realizes that the queue is about to be flooded by the big packets. This means, more possibilities to have less share of real-time packets, and congestion situation occurrence. Adhere; M–SQM perform early drop for the big FTP incoming packets. Then, involve ECN function to detect the congestive packets early stages, and remove FTP packets replaced by VoIP.

As more FTP flows imposed, the value of *msqm_thresh* then increases, M–SQM performs early aggressive dropping against most of the FTP packets. This is performed by, singing the threshold to smaller value. As a result most of incoming FTP packets will be in dropping procedure or enqueued with high dropping probability value. This behavior is performed till the queue equilibrium is perceived between FTP and VoIP traffic.

In Fig 9, the number of VoIP received packets is measured over the incremented value of big FTP flows. When the network is only dominated by fixed amount of VoIP flows, the loss rate will be zero, and no packet is dropped. So, all the schemes transmit almost the same number of VoIP packets.

As big FTP flows started to be introduced, and share the bandwidth with VoIP, the received VoIP traffic decreases a bit. M–SQM experienced this decrease at the initial phases. Although, more FTP flows are introduced, M–SQM starts to adopt itself so that favoring VoIP traffic. *msqm_thresh* is tuned to be preserved with relatively small value to promote those small VoIP packets.

By the time, the number of FTP flows exceeds the number of fixed VoIP flows "beyond 100 flows"; M–SQM cannot afford to keep the stable amount of favoring VoIP packets. This is because the queue is dominated by big FTP packets. At a certain point, where the value of *msqm_thresh* started to gradually increase far beyond the size of incoming packet, M–SQM uses an aggressive early drop mechanism that drops FTP packets.

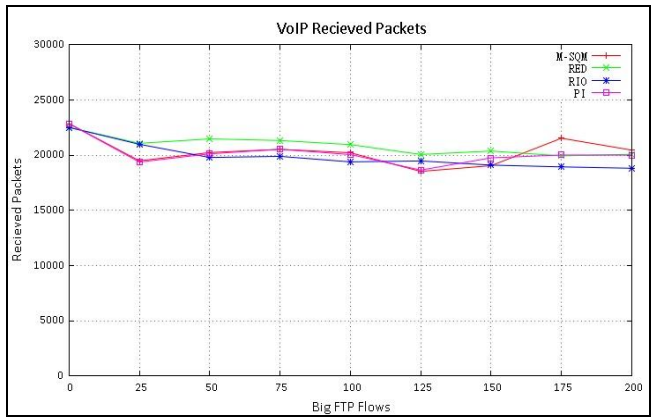

Fig. 9. VoIP received packets with respect to increased FTP flows.

This attempt is to decrease the value of *msqm_thresh* relatively small near to the value of VoIP packets. So that, more space is allocated for real-time packets. The action of dropping allows M–SQM to prioritize more VoIP packets over FTP, and outperform its peers at the peak points (175 and 200), whereby; the network is ultimately flooded with traffic.

### 5.4 Second Scenario: Total Link-delay with Varied FTP Traffic

In the plotted Fig. 10, in spite of the increased amount of the congestive big FTP traffic, M–SQM clearly outperforms the rest of AQM scheme. It records the lowest values of the link–delay. The aim here is to guarantee a low ratio of link delay against more congestive traffic.

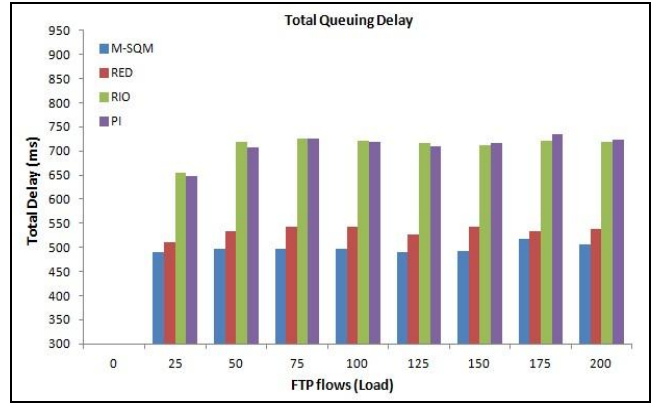

Fig. 10. Total-link delay with respect to increased FTP traffic.

As the flow numbers of big FTP increase, more FTP packets inter the queue. Therefore, it contributes to significant total link-delay. For each comparative algorithm, the queue is confined by the big FTP packets without imposing any strict policy to control the link-delay. Ironically, M–SQM scheme cares about guaranteeing low ratio of total queue-delay value in the congestive link. It rapidly detects the congestion caused by big flows, and applies suitable fair dropping policy against these incremented flows. This procedure guarantees bandwidth fair share, and level of prioritization to the real-time traffic.

As we mentioned early, RED still left a little performance to be desired in maintaining low link-delay in this scenario comparing to our M–SQM scheme. This is simply related to the low response of detecting the congestion situation. Wherein; RED requires controlling two thresholds, to keep the average queue size within the balanced range.

## 6 CONCLUSIONS

As a service differentiation scheme; M–SQM managed to provide fair classification for different traffic types, having their packet size as a criterion to the congestion level. However this prioritization seems to be in favor of VoIP traffic, but still there is a mechanism to control this prioritization to a certain level. This classification is to alleviate the congestion at the congestive link. ECN has this role to anticipate and handle the high traffic flooding. The simulation scenarios clearly show that M-SQM outperform the other AQMs in terms of received packets and total link-delay. It handles the congestion situations wisely in order to maintain fair share between VoIP and FTP. M-SQM finally is optimized to control the congestive traffic at the link between the gateways of the networks.

## 6 ACKNOWLEDGEMENT

THIS WORK WAS SUPPORTED BY THE MALAYSIAN MENISTRY OF HIGHER EDUCATION FOUNDATION RESEARCH GRANT SCHEME. FRGS NO: FRGS/1/11/GS/UPM/01/1.

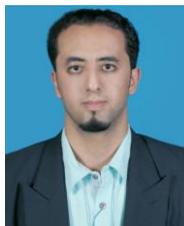

**Nadim K. M. Madi** received his Bachelor degree in Information Technology, majoring of Computer Networks from University Utara Malaysia, in 2010. He received his Master degree of Computer Sicence, Majoring of Distributed Computing from University Putra Malaysia, in 2012. Currently he is a .PhD. candidate in Computer Science in the same institution. His work includes network measurement and traffic analysis, Internet protocols and systems, Packet Scheduling Algorithms of Internet traffic, Internet QoS, fairness, congestion control of multiple traffic, wireless and mobile networking protocols. Up to 2012, he has published one paper in cloud computing and he was graduated with from the Master program with first-class honors place from university Putra Malaysia.

**Mohamed Othman** received his PhD from the National University of Malaysia with distinction (Best PhD Thesis in 2000 awarded by Sime Darby Malaysia and Malaysian Mathematical Science Society). Now, he is a Professor in Computer Science and Deputy Dean of Faculty of Computer Science and Information Technology, Universiti Putra Malaysia (UPM) and prior to that he was a Deputy Director of Information Development and Communication Center (iDEC) where he was incharge for UMPNet network campus, uSport Wireless Communication project and UPM DataCenter. In 2002 till 2010, he received many gold and silver medal awards for University Research and Development Exhibitions and Malaysia Technologies Exhibition which is at the national level. His main research interests are in the fields of parallel and distributed algorithms, high-speed networking, network design and management (network security, wireless and traffic monitoring) and scientific computing. He is a member of IEEE Computer Society, Malaysian National Computer Confederation and Malaysian Mathematical Society. He already published more than 150 National and International journals and more than 200 proceeding papers. He is also an associate researcher and coordinator of High Speed Machine at the Laboratory of Computational Science and Informatics, Institute of Mathematical Science (INSPEM), Universiti Putra Malaysia.